\documentstyle[psfig]{aipproc}

\def\la{\lesssim}
\def\ga{\gtrsim}
\def\bp{Paczy\'{n}ski}
\def\mesz{M\'{e}sz\'{a}ros}
\def\tE{t_{\oplus}}
\def\nuEm{\nu_{m}}
\def\FEm{F_{\nu, m}}
\def\FE{F_{\nu, \oplus}}
\def\nuE{\nu}

\begin{document}
\title{Afterglows as Diagnostics of Gamma Ray Burst Beaming}

\author{James E. Rhoads}
\address{Kitt Peak National Observatory\thanks{Kitt Peak 
National Observatory is part of
the National Optical Astronomy Observatories, operated by the
Association of Universities for Research in Astronomy.}\\
950 N. Cherry Ave.,
Tucson, AZ 85719
}

\maketitle

\begin{abstract}
If gamma ray bursts are highly collimated, radiating into only a small
fraction of the sky, the energy requirements of each event may be reduced
by several (up to $4$--$6$) orders of magnitude, and the event rate
increased correspondingly.   The large Lorentz factors ($\Gamma \ga 100$)
inferred from GRB spectra imply relativistic beaming of the gamma rays into
an angle $\sim 1/\Gamma$.  We are at present ignorant of whether there are
ejecta outside this narrow cone.

Afterglows allow empirical tests of whether GRBs are well-collimated
jets or spherical fireballs.  The bulk Lorentz factor decreases and
radiation is beamed into an ever increasing solid angle as the burst
remnant expands.  It follows that if gamma ray bursts are highly
collimated, many more optical and radio transients should be observed
without associated gamma rays than with them.  In addition, a burst
whose ejecta are beamed into angle $\zeta_m$ undergoes a qualitative
change in evolution when $\Gamma \zeta_m \la 1$: Before this, $\Gamma
\propto r^{-3/2}$, while afterwards, $\Gamma \propto
\exp(-r/r_{_\Gamma})$.  This change results in a potentially
observable break in the afterglow light curve.

Successful application of either test would eliminate the largest remaining
uncertainty in the energy requirements and space density of gamma ray
bursters.
\end{abstract}

The ejecta from gamma ray bursts must be highly relativistic to
explain the spectral properties of the emergent radiation
[1,4].  
The gamma rays we observe are
therefore only those from material moving within angle $1/\Gamma$ of
the line of sight, and offer no straightforward way of determining
whether the bursts are isotropic emitters or are beamed into a small
angle.  (Here $\Gamma$ is the bulk Lorentz factor of expansion.)

Afterglow emission at longer wavelengths is expected to arise later in
the evolution of the burst than the original gamma rays.  It therefore
offers at least two ways of testing the burst beaming hypothesis.

\subsection*{Burst and Afterglow Event Rates}
First, because $\Gamma$ is lower at the time of afterglow emission
than during the GRB itself, the afterglow cannot be as collimated as
the GRB can.  This implies that the afterglow event rate should
exceed the GRB event rate substantially if bursts are strongly
beamed.  Allowing for finite detection thresholds, 
\begin{equation} 
{N_{12} \over N_2 } \le {\Omega_1 \over \Omega_2 } \le {N_1 \over
N_{12}} ~~,
\label{jrhoads:ineq}
\end{equation}
where $N_1$, $N_2$ are the measured event rates above our detection
thresholds at our two frequencies; $N_{12}$ is the rate of events
above threshold at both frequencies; and $\Omega_1$, $\Omega_2$ are
the solid angles into which emission is beamed at the two frequencies.

A full derivation of this result and discussion of its application is
given in [6]. 
Rather than reproduce it, I
will refer the reader to that paper and will here discuss the second
test more fully than was possible in [6]. 

\subsection*{Dynamical Calculations: Numerical Integrations}
The second test is based on differences between the dynamical
evolution of beamed and isotropic bursts.  We explore the effects of
beaming on burst evolution using the notation
of~[5].  
Let $\Gamma_0$ and $M_0$ be the initial
Lorentz factor and ejecta mass, and $\zeta_m$ the opening angle into
which the ejecta move.  The burst energy is $E = \Gamma_0 M_0 c^2
\zeta_m^2 / 4$, where we assume a unipolar jet geometry.  Let $r$ be
the radial coordinate in the burster frame; $t$, $t_{co}$, and $\tE$
the time from the event measured in the burster frame, comoving ejecta
frame, and terrestrial observer's frame; and $f$ the ratio of swept up
mass to $M_0$.

The key assumptions in our beamed burst model are that (1) the energy and
mass per unit solid angle are constant at angles $\theta < \zeta_m$
from the jet axis and zero for $\theta > \zeta_m$ (see
[2] 
for an alternative model);
(2) the energy in the ejecta is
approximately conserved; (3) the ambient medium has uniform density;
and (4) the cloud of ejecta + swept-up material expands in its
comoving frame at the sound speed $c_s = c/ \sqrt{3}$ appropriate for
relativistic matter.  The last of these assumptions implies that the
working surface of the expanding remnant has a transverse size $\sim
\zeta_m r  + c_s t_{co}$.  The evolution of the burst changes when the
second term dominates over the first.

The full equations describing the burst remnant's evolution are then
\begin{equation}
f = {1\over M_0} \int_0^r r^2 \Omega_m(r) \rho(r) dr ~~,
\end{equation}
\begin{equation}
\Omega_m = \pi ( \zeta_m + c_s t_{co} / c t )^2
 \approx \pi ( \zeta_m +  t_{co} / \sqrt{3} t )^2 ~~,
\end{equation}
\begin{equation}
\Gamma = \left( \Gamma_0 + f \right) / \sqrt{ 1 + 2 \Gamma_0 f + f^2}
\approx \sqrt{ \Gamma_0 / 2 f } ~~,
\end{equation}
 %
\begin{equation}
 t = r/c~~, \qquad 
t_{co} = \int_0^t  dt' / \Gamma ~~, \quad \hbox{and} \quad
\tE =  \int_0^t  dt' / 2  \Gamma^2 ~~.
\end{equation}
 %

These equations can be solved by numerical integration to yield
$f(r)$, $\Gamma(r)$, and $\tE(r)$.  Figure~1 shows $\Gamma(r)$ from such
integrations for an illustrative pair of models (one beamed, one
isotropic).

\begin{figure}[h!] 
\centerline{\psfig{file=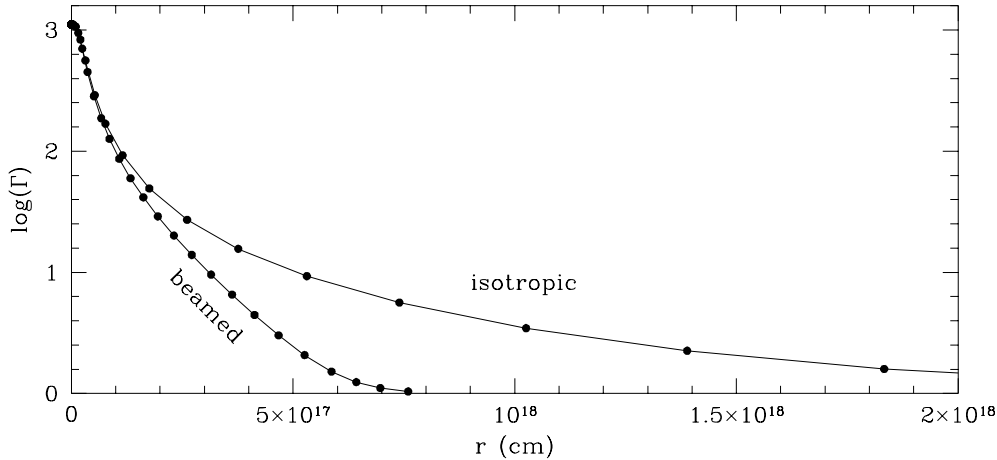,height=2.55in}} 
\vspace{10pt}
\caption{Dependence of the bulk Lorentz factor $\Gamma$ on the burst
expansion radius for an isotropic burst and a burst beamed into an
opening angle $\zeta_m = 0.01$ radian.  Both bursts follow a $\Gamma
\propto r^{-3/2}$ evolution initially, but the beamed burst changes its
behavior at
$\Gamma \approx 100 \approx 1/\zeta_m$, beyond which its Lorentz factor
decays exponentially with radius.
}
\label{jrhoads:fig1}
\end{figure}

The emergent synchrotron radiation can also be calculated if we assume
an electron energy spectrum and assume that  electrons and
magnetic fields have constant fractions of the equipartition energy density.
For illustrative purposes, we again follow the assumptions in
[5]. 
The electron energy spectrum is $N({\cal
E}) \propto {\cal E}^{-2}$, i.e. a power law with
equal energy per decade, so that the synchrotron spectrum peaks where $\tau =
0.35$, rising as $\nu^{5/2}$ at low (optically thick) frequencies and
falling as $\nu^{-1/2}$ at high (optically thin) frequencies
[3].  
The relevant equations are a straightforward modification of equations
11--20 of [5]. 
Figure~2 shows the peak flux
density as a function of observed frequency for the models used in figure~1.
We caution the reader that more recent 
electron energy spectra grounded in observations
(e.g. [7])  
may be more reliable.  We hope to
incorporate such spectra in our calculations in future.

\begin{figure}[ht!] 
\centerline{\psfig{file=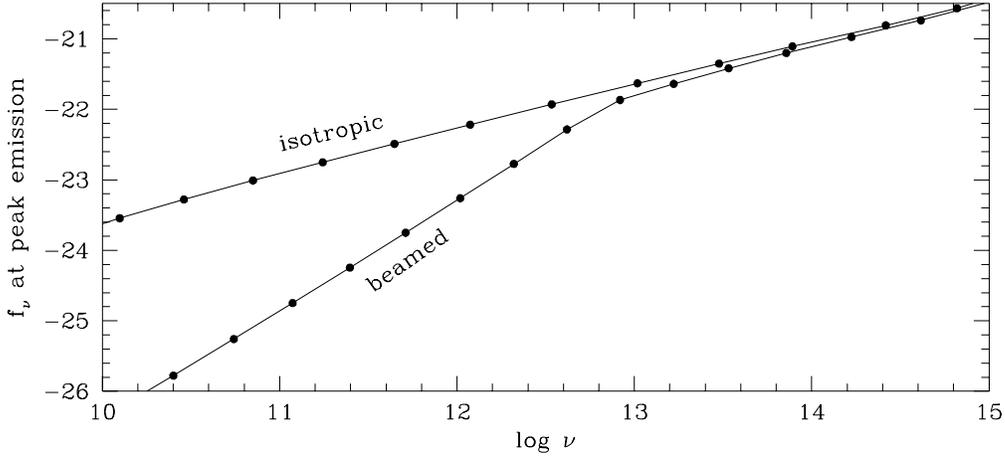,height=2.55in}} 
\vspace{10pt}
\caption{The dependence of the peak flux density $f_\nu$ on observed
frequency $\nu$ for the same pair of bursts.  The electron spectrum
follows the model of \bp\ \& Rhoads (1993).  The peak in the
synchrotron emission for this model occurs at the frequency where
optical depth effects become important. 
The predicted break in the power law caused by beaming should be
observable.  Similar breaks occur in the dependence of $f_\nu$ and
$\nu_{peak}$ with time, and are expected to be a generic feature of 
beamed GRB afterglow models.
}
\label{jrhoads:fig2}
\end{figure}

\subsection*{Dynamical Calculations: Analytic Integrations}
The most interesting dynamical change introduced by beaming is a
transition from a power law $\Gamma \propto r^{-3/2}$ to an exponentially
decaying regime $\Gamma \propto  \exp(-r/r_{_\Gamma})$. 
This can be derived by considering the approximate evolution equations
for the regime where (a) $1/\Gamma_0 \la f \la \Gamma_0$, so that
$\Gamma \approx \sqrt{\Gamma_0 / 2 f }$;  and (b) 
$c_s t_{co} > \zeta_m r$ (corresponding to $f \ga 9
\Gamma_0 \zeta_m^2$):
%
\begin{equation}
d f / d r \approx {\pi \over M_0} c_s^2 t_{co}^2 \rho \quad , \quad
d t_{co} / d r  \approx \sqrt{2 f \over c^2 \Gamma_0} \quad , \quad
d \tE / d r \approx { f \over c \Gamma_0 } \quad .
\label{jrhoads:eqn_simpler}
\end{equation}
It follows that 
\begin{equation}
\sqrt{f} d f = {\pi \over \sqrt{2}} { c\, c_s^2 \rho \sqrt{\Gamma_0}
\over M_0} \times  t_{co}^2 d t_{co}
 \approx  {\pi \over 3 \sqrt{2}} { c^3 \rho \sqrt{\Gamma_0}
\over M_0} \times t_{co}^2 d t_{co} ~~.
\end{equation}
This is easily integrated to obtain
\begin{equation}
f^{3/2} = \left( \pi \sqrt{\Gamma_0} c\, c_s^2 \rho \over \sqrt{8}  M_0
\right) t_{co}^3 + \hbox{\sl const} ~~.
\label{jrhoads:eqn_f_tco}
\end{equation}
The constant of integration becomes
negligible once $c_s t_{co} \gg \zeta_m r$, so that
equation~\ref{jrhoads:eqn_f_tco} becomes $f \propto t_{co}^2$.  It is
then clear from equations~\ref{jrhoads:eqn_simpler} that $f$,
$\Gamma$, $t_{co}$, and $\tE$ will all behave exponentially with $r$ in
this regime.  Retaining the constants of proportionality, we find
\begin{equation}
f \propto \exp( 2 r / r_{_\Gamma} ) \qquad \hbox{where} \qquad
r_{_\Gamma} = \left[ {1 \over \pi} \left(c \over c_s \right)^2 
{ \Gamma_0 M_0 \over \rho } \right]^{1/3} ~~.
\end{equation}
Further algebra yields $\Gamma \propto \exp(-r / r_{_\Gamma} )$ and 
$\tE \propto f \propto \exp( 2 r /  r_{_\Gamma} )$,
so that $\Gamma \propto \tE^{-1/2}$.
Thus, while the evolution of $\Gamma(r)$ changes from a power law to an
exponential at $\Gamma \sim 1/\zeta_m$, the evolution of $\tE(r)$
changes similarly.  The net result is that $\Gamma(\tE)$ has a power
law form in both regimes, but with a break in the slope from $\Gamma
\propto \tE^{-3/8}$ when $\Gamma >  1/\zeta_m$ to $\Gamma \propto
\tE^{-1/2}$ when $\Gamma < 1/\zeta_m$.

Of course, $\Gamma$ is not directly observable, and we ultimately want
to predict 
observables like the frequency of peak emission $\nuEm$, the flux
density $\FEm$ at $\nuEm$, and the angular size $\theta$ of the afterglow.
With the electron energy spectrum
described above, the relevant power law scalings before beaming
becomes dynamically important are $\nuEm \sim \tE^{-2/3}$,  $\FEm \sim
\tE^{-5/12}$, and  $\theta \sim \tE^{5/8}$.  At late times, 
$\nuEm \sim \tE^{-1}$, $\FEm \sim \tE^{-3/2}$, and $\theta \sim \tE^{1/2}$.
%
%
Our numerical integrations confirm these relations, though the transition
between the two regimes is quite gradual for $\nuEm$.

Combining these scalings with the spectral shape
yields predictions for the light curve at fixed observed
frequency.  The most dramatic feature is in the light curve shape for
$\nuE > \nuEm$, which changes from $\FE \sim \tE^{-3/4}$ to $\FE \sim
\tE^{-2}$.  These exponents are generally sensitive to the assumed
electron energy distribution in the blast wave.

\subsection*{Conclusions}
Establishing whether or not gamma ray bursts are beamed will be 
valuable in understanding source populations and burst mechanisms.
There are at least two potentially observable consequences of beaming.

(1) The event rate for afterglows should exceed that for bursts
substantially if bursts are strongly beamed.  A quantitative
comparison of rates at two frequencies yields quantitative limits on
the ratio of beaming angles.

(2) The dynamical evolution of a beamed burst remnant changes
qualitatively when $\Gamma < 1/\zeta_m$.  The resulting changes in the
light curves could be observed.

\end{document}